# The heavy quark decomposition of the S-matrix and its relation to the pinch technique.

Joannis Papavassiliou, Kostas Philippides

and

Martin Schaden

Department of Physics, New York University, 4 Washington Place,

New York, NY 10003, USA.

## ABSTRACT

We propose a decomposition of the S-matrix into individually gauge invariant sub-amplitudes, which are kinematically akin to propagators, vertices, boxes, etc. This decompsition is obtained by considering limits of the S-matrix when some or all of the external particles have masses larger than any other physical scale. We show at the one-loop level that the effective gluon self-energy so defined is physically equivalent to the corresponding gauge independent self-energy obtained in the framework of the pinch technique. The generalization of this procedure to arbitrary gluonic $n$-point functions is briefly discussed.

The pinch technique (PT) [1] is an algorithm that allows the construction of modified gauge invariant (g.i.) n-point functions, through the order by order rearrangement of the Feynman graphs contributing to a certain physical, and therefore ostensibly g.i. amplitude (such as an S-matrix element, or a Wilson loop). Even though the most recent applications of the PT are inspired by Standard Model phenomenology [2-7], it was originally introduced in the context of QCD, as a first step toward the construction of Schwinger-Dyson equations, which would respect the crucial property of gauge invariance, even in their one-loop dressed truncated version [8-9]. The simplest example that demonstrates how the PT works is the gluon two point function (propagator). Consider the S-matrix element $T$ for an elastic scattering process such as $q_1 \bar{q}_2 \to q_1 \bar{q}_2$, where $q_1,q_2$ are two on-shell test quarks with masses $m_1$ and $m_2$. To any order in perturbation theory $T$ is independent of the gauge fixing parameter $\xi$. On the other hand, as an explicit calculation shows, the conventionally defined proper self-energy [collectively depicted in graph 1(a)] depends on $\xi$. At the one loop level this dependence is canceled by contributions from other graphs, such as 1(b), 1(c), 1(d), and 1(e) which, at first glance, do not seem to be propagator-like. That this cancellation must occur and can be employed to define a g.i. self-energy, is evident from the decomposition:

$$T(s,t,m_1,m_2) = T_0(t,\xi) + T_1(t,m_1,\xi) + T_2(t,m_2,\xi) + T_3(s,t,m_1,m_2,\xi) , \qquad (1)$$

where the function $T_0(t,\xi)$ depends kinematically only on the Mandelstam variable $t = -(\hat{p}_1 - p_1)^2 = -q^2$, and not on $s = (p_1 + p_2)^2$ or on the external masses. Typically, self-energy, vertex, and box diagrams contribute to $T_0$, $T_1$ and $T_2$, and $T_3$, respectively. Such contributions are $\xi$ dependent, in general. However, as the sum $T(s,t,m_1,m_2)$ is g.i., it is easy to show that Eq. (1) can be recast in the form

$$T(s,t,m_1,m_2) = \hat{T}_0(t) + \hat{T}_1(t,m_1) + \hat{T}_2(t,m_2) + \hat{T}_3(s,t,m_1,m_2) , \qquad (2)$$

where the $\hat{T}_i$ ($i = 0,1,2,3$) are *individually* $\xi$-independent. The propagator-like parts 1(f), 1(g), 1(h), and 1(i), stemming from graphs 1(b), 1(c), 1(d), and 1(e), respectively, enforce the gauge independence of $T_0(t)$, and are called "pinch parts". They emerge every time a gluon propagator or an elementary three-gluon vertex contributes a longitudinal $k_\mu$ to the original graph's numerator. The action of such a term is to trigger an elementary Ward identity of the form $\not{k} = (\not{p} + \not{k} - m) - (\not{p} - m)$ when it gets contracted with a $\gamma$ matrix. The first term removes the internal fermion propagator (e.g. it produces a "pinch"), whereas the second vanishes on shell. From the g.i. functions $\hat{T}_i$ ($i = 1, 2, 3$) one may now extract a g.i. effective gluon ($G$) self-energy $\hat{\Pi}_{\mu\nu}(q)$, g.i. $Gq_i\bar{q}_i$ vertices $\hat{\Gamma}_\mu^{(i)}$, and a g.i. box $\hat{B}$, in the following way:

$$\begin{aligned}
\hat{T}_0 &= g^2 \bar{u}_1 \gamma^\mu u_1 [(\frac{1}{q^2})\hat{\Pi}_{\mu\nu}(q)(\frac{1}{q^2})] \bar{u}_2 \gamma^\nu u_2 \\
\hat{T}_1 &= g^2 \bar{u}_1 \hat{\Gamma}_\nu^{(1)} u_1 (\frac{1}{q^2}) \bar{u}_2 \gamma^\nu u_2 \\
\hat{T}_2 &= g^2 \bar{u}_1 \gamma^\mu u_1 (\frac{1}{q^2}) \bar{u}_2 \hat{\Gamma}_\nu^{(2)} u_2 \\
\hat{T}_3 &= \hat{B}
\end{aligned} \quad (3)$$

where $u_i$ are the external spinors, and $g$ is the gauge coupling. Since all hatted quantities in the above formula are g.i., their explicit form may be calculated using any value of the gauge-fixing parameter $\xi$, as long as one properly identifies and allots all relevant pinch contributions. The choice $\xi = 1$ simplifies the calculations significantly, since it eliminates the longitudinal part of the gluon propagator. Therefore, for $\xi = 1$ the pinch contributions originate only from momenta carried by the elementary three-gluon vertex of graph 1(b) (and its mirror graph, not shown). The one-loop expressions of $\hat{\Pi}_{\mu\nu}(q)$ and $\hat{\Gamma}_\mu^{(i)}$ are given by [2], [9]:

$$\hat{\Pi}_{\mu\nu}(q) = \Pi_{\mu\nu}^{(\xi=1)}(q) + t_{\mu\nu} \Pi^P(q) \quad (4)$$

with $t_{\mu\nu} = q^2 g_{\mu\nu} - q_\mu q_\nu$, and

$$\begin{aligned}
\Pi^P(q) &= -2ic_a g^2 \int_n \frac{1}{k^2(k+q)^2} \\
&= 2c_a (\frac{g^2}{16\pi^2})[C_{UV} - \ln(\frac{-q^2}{\mu^2}) + 2]
\end{aligned} \quad (5)$$

where $C_{UV} = \frac{2}{\epsilon} - \gamma + \ln(4\pi)$, $\epsilon = n - 4$, $\gamma = 0.577...$ is the Euler constant, $\int_n \equiv \int \frac{d^n k}{(2\pi)^n}$ is the dimensionally regularized loop integral. Similarly, for the vertex we have:

$$[\hat{\Gamma}_\mu]^{(i)} = ig^2 \left[ (\frac{c_a}{2}) \int_n \frac{\gamma^\rho S_i(p_i - k)\gamma^\sigma \Gamma^F_{\rho\sigma\mu}}{k^2(k+q)^2} + (\frac{c_a}{2} - c_f) \int_n \frac{\gamma^\rho S_i(\hat{p}_i - k)\gamma_\mu S_i(p_i - k)\gamma_\rho}{k^2} \right]$$
$$+ i \left[ \gamma^\mu \frac{1}{\slashed{p}_i - m_i} \hat{\Sigma}_i(p) + \hat{\Sigma}_i(\hat{p}_i) \frac{1}{\slashed{\hat{p}}_i - m_i} \gamma^\mu \right] \quad (6)$$

where $\Gamma^F_{\mu\nu\alpha} = 2q_\mu g_{\nu\alpha} - 2q_\nu g_{\nu\alpha} - (2k+q)_\alpha g_{\mu\nu}$ [10], $c_f$ is the Casimir eigenvalue of the fermion representation, $\hat{p}_i = p_i + q$, and

$$\hat{\Sigma}_i(p) = g^2 c_f \int_n \frac{1}{k^2} \gamma_\sigma \frac{1}{\slashed{k} + \slashed{p} - m_i} \gamma^\sigma = \Sigma_i^{(\xi=1)}(p) \quad (7)$$

is the one loop g.i. quark propagator, derived in [11]

In principle, this procedure can be generalized to an arbitrary $n$-point function. In particular, the g.i. three and four point functions $\hat{\Gamma}_{\mu\nu\alpha}$ and $\hat{\Gamma}_{\mu\nu\alpha\beta}$ have been derived in [9] and [12] The Green's functions obtained via the PT, in addition to being g.i., are endowed with several characteristic properties. Most noticeably, the gluon $n$-point functions computed thus far $[\hat{\Pi}_{\mu\nu} = t_{\mu\nu}\hat{\Pi}, \hat{\Gamma}_{\mu\nu\alpha}, \hat{\Gamma}_{\mu\nu\alpha\beta}$ $(n = 2, 3, 4)]$ satisfy the following simple QED-like Ward identities:

$$q_1^\mu \hat{\Gamma}_{\mu\nu\alpha}(q_1, q_2, q_3) = t_{\nu\alpha}(q_2)\hat{d}^{-1}(q_2) - t_{\nu\alpha}(q_3)\hat{d}^{-1}(q_3)$$
$$q_1^\mu \hat{\Gamma}^{abcd}_{\mu\nu\alpha\beta} = f_{abp}\hat{\Gamma}^{cdp}_{\nu\alpha\beta}(q_1 + q_2, q_3, q_4) + c.p. , \quad (8)$$

where $\hat{d}^{-1}(q) = q^2 - \hat{\Pi}(q)$, $f^{abc}$ are the structure constants of the gauge group, and the abbreviation c.p. in the rhs stands for "cyclic permutations". In addition, the gluon-quark vertices $[\hat{\Gamma}^a_\mu]^{(i)}$ of Eq. (6) are ultra-violet finite.

Regardless of any such properties, however, an ambiguity is associated with the construction of Green's functions via the PT. It is obvious for instance that, after a g.i. gluon self-energy and gluon-quark vertex has been constructed via the PT, one still has the freedom to add an arbitrary term of the form $(q^2 g_{\mu\nu} - q_\mu q_\nu)f(q^2)$ to the self-energy, and subtract it from the vertex. As long as the function $f(q^2)$ is g.i., such an operation satisfies the criterion of individual gauge invariance for the self-energy and vertex, respects their Ward identities, and preserves the uniqueness of the S-matrix. It is therefore desirable to have a physical prescription which eliminates this ambiguity. In this paper we propose an alternative, physically motivated prescription for extracting g.i. sub-amplitudes of the S-matrix, that are kinematically akin to self-energies, vertices and boxes. In this framework, the effective one-loop gluon self-energy is defined to be the limit of the S-matrix as both external fermion masses $m_1$ and $m_2$ are taken to be larger than any other mass scale in the process (they are however comparable to each other, e.g. $m_1 \approx m_2$). To one loop we show that this limit coincides with the static quark-antiquark potential for very massive quarks [13-14], and is physically equivalent to the PT result of Eq. (5).

Any reasonable effective propagator should only depend on the momentum transfer $t$ but not on kinematical details such as masses or total momentum $s$ of the incoming or outgoing particles. Similarly, any viable definition of a $Gq_i\bar{q}_i$ vertex should only depend on $t$ and the quark mass $m_i$ [15] and no other kinematical details. This reasoning can obviously be generalized to higher $n$-point functions. Motivated by these observations, we propose to define propagator and vertex-like sub-amplitudes by taking appropriate kinematical limits of the S-matrix.

For the simple case of a four-quark on shell amplitude $T$ (Fig.1) we define the following three limits

$$L_0(t) = T(s, t, m_1 = M, m_2 = M)$$
$$L_1(t, m_1) = T(s, t, m_1, m_2 = M) \qquad (9)$$
$$L_2(t, m_2) = T(s, t, m_1 = M, m_2)$$

where the mass $M$ is assumed to be larger than any other mass scale appearing in the process, except for any cutoffs introduced in intermediate calculations in order to regularize ultra-violet divergences. Note however that, since the external particles are on shell, $s = (p_1 + p_2)^2 \geq (m_1 + m_2)^2$ is also of the order of $M^2$, in any of these limits [16]. Each of the above quantities is g.i., since it corresponds to a particular limit of the g.i. S-matrix element $T$. They can be systematically computed by expanding the S-matrix in powers of $(\frac{\mu_0}{M})$, where $\mu_0$ is any of the remaining mass scales. The limits considered above correspond to well-defined physical situations. $L_0$, for example, is the dominant contribution to the S-matrix when the momentum transfer $t$ is considerably smaller than the masses of all the scattered particles, e.g. $t = -q^2 \ll m_1, m_2$.

We can define the following linear combinations:

$$\tilde{T}_0(t) = L_0$$
$$\tilde{T}_i(t, m_i) = (L_i - L_0) \quad (i = 1, 2) \tag{10}$$
$$\tilde{T}_3(s, t, m_1, m_2) = T(s, t, m_1, m_2) - L_0 - [(L_1 - L_0) + (L_2 - L_0)]$$

We have thus arrived at a decomposition of the S-matrix into individually g.i. and kinematically distinct sub-amplitudes, which we can identify as effective self-energy $\tilde{T}_0(t)$, vertices $\tilde{T}_1(t, m_1)$ and $\tilde{T}_2(t, m_2)$, and boxes $\tilde{T}_3(s, t, m_1, m_2)$. Clearly, the sum of these sub-amplitudes is the original S-matrix, e.g.

$$T(s, t, m_1, m_2) = \tilde{T}_0(t) + \tilde{T}_1(t, m_1) + \tilde{T}_2(t, m_2) + \tilde{T}_3(s, t, m_1, m_2) , \tag{11}$$

The above decomposition of the S-matrix into individually g.i. and kinematically distinct sub-amplitudes, relies on a procedure different from the PT. The question that naturally arises is how the individual terms of Eq. (2) and Eq. (11) are related. As we will show by an explicit one-loop calculation, $\hat{T}_0$ of Eq. (2) and $\tilde{T}_0$ of Eq. (10) are related as follows:

$$\tilde{T}_0(t) = \hat{T}_0(t) + g^2 \bar{u}_1 \gamma^\mu u_1 (\frac{1}{q^2})[C t_{\mu\nu}](\frac{1}{q^2}) \bar{u}_2 \gamma^\nu u_2 . \tag{12}$$

In Eq. (12) $C$ is a g.i. finite numerical constant. Thus the g.i. self-energy $\tilde{\Pi}_{\mu\nu}(q)$ extracted from $\tilde{T}_0(t)$, and the $\hat{T}_0$ obtained form Eq. (2) satisfy:

$$\tilde{\Pi}_{\mu\nu}(q) = \hat{\Pi}_{\mu\nu}(q) + C t_{\mu\nu} \tag{13}$$

Clearly, the term proportional to $C$ in the r.h.s. of Eq. (13) can be removed by a finite counterterm, or, equivalently, absorbed in the final normalization of the S-matrix.

To compute the leading one-loop contribution to $\tilde{T}_0(t)$ (or equivalently $L_0$) we evaluate the S-matrix in the limit $m_1 \approx m_2 \to M$, where $M \gg -q^2$. [17] For simplicity we consider elastic scattering, so that $q^2 < 0$. As a consequence, there are no imaginary parts in the Feynman graphs. We define the Euclidean momentum $Q^2 = -q^2 > 0$. Throughout the calculation we use dimensional regularization, where the UV cutoff is set by the usual pole $\frac{1}{\epsilon}$. In addition, the t'Hooft mass $\mu$ has to be introduced. The infrared divergences are regulated by introducing an infrared gluon mass $\lambda$ in the intermediate calculations. [18]

We then compute all one-loop Feynman graphs contributing to the process, neglecting terms proportional to any of the ratios $(\frac{Q}{M})$, $(\frac{\lambda}{M})$, and $(\frac{\mu}{M})$, (or higher powers of such ratios), and retaining only logarithmic and constant terms. We emphasize that the above expansion is carried out *after* the integration over the loop momenta has been performed in dimensional regularization. Effectively this means that $M$ is always much smaller than the cutoff $\Lambda$ [e.g. $\frac{1}{\epsilon} \to \ln(\frac{\Lambda}{\mu}) \gg \ln(\frac{M}{\mu})$]. In this calculation all choices for $\xi$ are equivalent, since the S-matrix element is $\xi$-independent; we choose $\xi = 1$ for convenience.

The most involved part of the calculation are the box diagrams. It is important to recognize that both the direct and the crossed graph must be appropriately combined in order to obtain the correct color structure. It is also interesting to notice that the expressions that survive the large $M$ limit are of non-Abelian nature only, namely proportional to $c_a$. If we call $B_{dir}$ the total contribution of the direct graph and $B_{cr}$ the respective contribution from the crossed, we have that $B_{dir} = (R_a R_b)_1 (R_a R_b)_2 S_{dir}$ and $B_{cr} = (R_a R_b)_1 (R_b R_a)_2 S_{cr}$ where $S_{dir}$ and $S_{cr}$ are the remainders of the boxes, after the color structure has been factored out. The important step is to show that in the large $M$ limit we have $S_{dir} = -S_{cr}$. Thus, the total box contribution $\tilde{B}$ becomes:

$$\tilde{B} = (R_a R_b)_1 [R_a, R_b]_2 S_{dir}$$
$$= \frac{1}{2} c_a (R_c)_1 (R_c)_2 S_{dir} \quad (14)$$

The result for the individual Feynman graphs are: (we omit external spinors and an overall factor of $\frac{g^2}{Q^2}$)

$$[(a)] = \Pi_{\mu\nu}^{(\xi=1)}$$

$$[(b) + (b)_{mirror}] = \frac{g^2}{16\pi^2} c_a \left[ 3C_{UV} + 4 + 3\ln(\frac{\mu^2}{M^2}) \right] g_{\mu\nu} + ...$$

$$[(c) + (c)_{mirror}] = \frac{g^2}{16\pi^2}(2c_f - c_a) \left[ C_{UV} + 4 + 2\ln(\frac{\lambda^2}{\mu^2}) + 3\ln(\frac{\mu^2}{M^2}) \right] g_{\mu\nu} + ... \quad (15)$$

$$[(d) + (d)_{mirror}] = -2\frac{g^2}{16\pi^2} c_f \left[ C_{UV} + 4 + 2\ln(\frac{\lambda^2}{\mu^2}) + 3\ln(\frac{\mu^2}{M^2}) \right] g_{\mu\nu} + ...$$

$$[(e) + (e)_{crossed})] = 2\frac{g^2}{16\pi^2} c_a \ln(\frac{\lambda^2}{Q^2}) g_{\mu\nu} + ...$$

where the ellipsis denote terms of order $O(\frac{1}{M})$ or higher. We notice that the sum of all vertex and box graphs listed in Eq. (15) is equal to $2\frac{g^2 c_a}{16\pi^2}[C_{UV} - \ln(\frac{Q^2}{\mu^2})]$, which is, up to a physically irrelevant constant, the pinch contribution to the self-energy, given in Eq. (5). The total contribution to $\tilde{T}_0$ reads:

$$\tilde{T}_0 = g^2 \bar{u}_1 \gamma^\mu u_1 (\frac{1}{q^2}) \left[ \Pi_{\mu\nu}^{(\xi=1)} + 2(\frac{g^2 c_a}{16\pi^2})[C_{UV} - \ln(\frac{-q^2}{\mu^2})] t_{\mu\nu} (\frac{1}{q^2}) \right] \bar{u}_2 \gamma^\nu u_2$$

$$= g^2 \bar{u}_1 \gamma^\mu u_1 (\frac{1}{q^2}) \left[ \Pi_{\mu\nu}^{(\xi=1)} + \Pi^P(q) - 4(\frac{g^2 c_a}{16\pi^2}) t_{\mu\nu} \right] (\frac{1}{q^2}) \bar{u}_2 \gamma^\nu u_2 \quad (16)$$

$$= \hat{T}_0 - g^2 \bar{u}_1 \gamma^\mu u_1 (\frac{1}{q^2})[4(\frac{g^2 c_a}{16\pi^2}) t_{\mu\nu}](\frac{1}{q^2}) \bar{u}_2 \gamma^\nu u_2$$

which is the advertised result in the first line of Eq. (12), with $C = -4(\frac{g^2 c_a}{16\pi^2})$. The first relation of Eq. (13) follows immediately from Eq. (16), namely

$$\tilde{\Pi}_{\mu\nu}(q) = \hat{\Pi}_{\mu\nu}(q) - 4(\frac{g^2 c_a}{16\pi^2}) t_{\mu\nu} \quad (17)$$

Adding the tree-level contribution to $\tilde{T}_0$ of Eq. (16), and using the standard result

$$\Pi_{\mu\nu}^{(\xi=1)} = \frac{g^2 c_a}{16\pi^2} \left[ \frac{5}{3}[C_{UV} - \ln(\frac{-q^2}{\mu^2})] + \frac{31}{9} \right] t_{\mu\nu} \quad (18)$$

together with Eq. (5), we find that $\tilde{T}_0$ is identical to the Fourier transform of the unrenormalized one-loop static potential $V(Q^2)$ for a heavy quark-antiquark system ( [13]), namely (we omit the external spinors),

$$\tilde{T}_0 = V(Q^2) = -\frac{g^2 c_a}{Q^2}\left[1 + \frac{g^2 c_a}{16\pi^2}\{-\frac{11}{3}C_{UV} + \frac{11}{3}\ln(\frac{Q^2}{\mu^2}) + \frac{31}{9}\}\right] \qquad (19)$$

where the factor $(\frac{11}{3})\frac{c_a}{16\pi^2}$ is the coefficient $b_0$ in front of $-g^3$ in the one loop $\beta$ function. The contribution $\tilde{T}_0$ of Eq. (16) or Eq. (19) to the S-matrix is infrared finite. The decomposition of Eq. (9), together with Eq. (16), implies that the PT result for the self-energy gives the dominant contribution to the physical S-matrix, when the scattered particles are heavy compared to other mass scales. We have thus arrived at a physical interpretation of this PT sub-amplitude. In that sense, the mathematical ambiguity in defining a g.i. propagator-like sub-amplitude of the S-matrix, which we discussed previously, can be eliminated by imposing a physically motivated boundary condition, i.e. that the effective self-energy should reproduce the S-matrix for the scattering of sufficiently heavy external quarks. Since perturbation theory in QCD is reliable only for momentum transfers beyond a few GeV, in practice this sub-amplitude will provide the dominant contribution to the S-matrix only for top and bottom scattering. Nevertheless, it makes sense to *define* this g.i. sub-amplitude also for considerably lighter systems, although in such a case it will generally not give the dominant contribution to the S-matrix.

It would clearly be of interest to extend this analysis to the vertex-like sub-amplitudes. Of course, in a theory with massless gauge bosons such sub-amplitudes are in general infrared divergent; they can therefore not be directly related to a physical process, without including bremsstrahlung. One could nevertheless compare the g.i. vertex-like amplitudes $\hat{T}_i$ and $\tilde{T}_i$, $i = 1, 2$, of the two schemes, as long as the infra-red singularities are regulated in a gauge invariant manner, such as dimensional infra-red regularization [19-20]. This goes however beyond the scope of the present communication.

The previous considerations can be generalized to the case of multi-quark scattering. In particular, from a $2n$-quark amplitude one can define a g.i. gluon $n$-point functions $\tilde{\Gamma}^{(n)}(q_1, ..., q_n)$ with all incoming momenta $q_i$, $i = 1, ..n$ off-shell. To that end one has to consider the limit of the amplitude as all external fermion masses become large ($m_i \to M$, $i = 1, .., n$). It would be very interesting to determine if the g.i. $n$-point functions so obtained are physically equivalent to those obtained with the PT, especially for $n = 3, 4$. It would be also interesting to generalize the previous arguments to the case of theories with spontaneous symmetry breaking in general, and the electro-weak sector of the Standard Model, in particular.

## 1. Acknowledgment

The authors thank A. Sirlin for helpful discussions. This work was supported by the National Science Foundation under Grant No.PHY-9017585. K.P. acknowledges support from the E.C. network ERB-CHRXT 930319. The work of M.S. was supported by Deutsche Forschungsgemeinschaft under Grant No.Scha/1-1.

17. It is important to notice that, since the external quarks are "on shell" the incoming momenta $p_i$ are not to be neglected compared to $M$ Therefore, our results will in general be different from the case where all incoming momenta are small compared to a large mass inside the loop.

18. Alternatively, one could use dimensional regularization for both ultra-violet and infra-red divergences, as in Ref. [13].

19. R. Gastmans and R. Meuldermans, Nucl. Phys. B 63, 277 (1973).

20. W. J. Marciano and A. Sirlin, Nucl. Phys. B 88, 86 (1975).

## 3. Figure Caption

Graphs (a)-(e) are some of the QCD contributions to the S-matrix for four-fermion processes. Graphs (f), (g), (h), and (i) are the pinch parts of (b), (c), (d), and (e) respectively. When added to the usual self-energy graphs (a), they give rise to the $\xi$-independent amplitude $\hat{T}_0(t)$. The mirror image graphs corresponding to (b), (c) and (d), as well as the crossed box graph are not shown. Graph (a) contains contributions from virtual fermions, gluons and Faddeev-Popov ghosts. Solid (wavy) lines represent fermions (gluons).